\documentclass{aa}
\usepackage{graphicx}

\newcommand{\Teff}  {\mbox{T$_\mathrm{eff}$}\,}
\newcommand{\FeH}   {\mbox{[Fe/H]\,}}
\newcommand{\logg}  {\mbox{$\log$ g}\,}
\newcommand{\tgm}   {\mbox{(\Teff, \logg, [Fe/H])} \,}

\begin{document}

\title{Catalogue of \FeH determinations for FGK stars : 
2001 edition\thanks{The catalogue (Tables 1 and 2) is only available in electronic form at the Centre de Donn\'ees Stellaires in
Strasbourg  (http://cdsweb.u-strasbg.fr/Cats.html) and VizieR (http://vizier.u-strasbg.fr/).}
}

\author{
G. Cayrel de Strobel \inst{1} \and C. Soubiran \inst{2} 
\and N. Ralite \inst{2}}

\offprints{Caroline.Soubiran@observ.u-bordeaux.fr}

\institute{
Observatoire de Paris-Meudon, CNRS UMR 8633, 92195 Meudon Cedex, France
\and 
Observatoire de Bordeaux, CNRS UMR 5804, BP 89, 33270 Floirac, France}

\date{Received 20 Febuary 2001; accepted 19 March 2001}

\titlerunning{[Fe/H] catalogue}

\abstract{
The catalogue presented here is a compilation of published atmospheric 
parameters \tgm \,
obtained from high resolution, high signal-to-noise spectroscopic observations.
This new edition
has changed compared to the five previous versions. It is now restricted to
intermediate and low mass stars (F, G and K stars).
It contains 6354 determinations of \tgm
for 3356 stars, including 909 stars in 79 stellar systems. 
The literature is 
complete between January 1980 and December 2000 and includes 378 references. The
catalogue is made up of two tables, one for field stars and one for
stars in galactic associations, open and globular clusters and external galaxies. 
The catalogue is distributed through the CDS database.
Access to the catalogue with cross-identification to other sets of data is also possible with VizieR
 (Ochsenbein et al \cite{och00}).
}
   \maketitle



\section{Introduction}

The \FeH catalogue is an exhaustive compilation of references presenting determinations 
obtained by detailed analyses of
the stellar atmospheric parameters \tgm relying on high resolution, 
high signal-to-noise spectroscopic observations. Such observations have enabled the accurate
measurements of equivalent width of weak metallic lines, which are proportional to the
abundances of the corresponding elements.

 Drastic changes have been introduced in the 2001 edition of the
\FeH catalogue as compared to the five previous ones (Cayrel de Strobel et al
\cite{cay80}, \cite{cay81}, \cite{cay85}, \cite{cay92}, \cite{cay97}). The  first change 
concerns the removal of stars hotter than 7000 K, the second concerns 
the removal of references prior to 1980  (mostly based on photographic material). The 1996 version 
(Cayrel de Strobel et al \cite{cay97}), which supersedes the older ones,
is the work of reference for \FeH determinations prior to 1980. 

The \FeH catalogue is now particularly suited for
studies of chemical evolution by means of middle and low mass stars. Indeed, the 
 abundances of such stars reflect, at least approximately, the chemical composition 
of the interstellar medium
 out of which they were formed. Massive stars 
 have their atmospheric abundances modified by internal structure processes producing a large 
set of chemical peculiarities. The transition between low and high
 mass stars has been taken at about \Teff = 7000 K.

 The chemical evolution of the Galaxy is assessed using 
the  metal/hydrogen ratios in stars.
Nowadays, specialists of galactic evolution are also very interested in the behaviour of the 
abundances
of C, N, O, Mg, Si... in stars belonging to different populations. But the small number of 
lines of 
these species
in the observed spectra results in iron being still the most widly used metallicity parameter. 
Unlike carbon, oxygen and magnesium, iron lines are extremely numerous in optical and 
near UV spectra.

A very important step in a spectral analysis of a star is the determination of accurate
atmospheric parameters for the selection of the appropriate model atmosphere. No abundance can 
be
derived unless the three physical parameters, effective temperature (\Teff), surface gravity (g) and
microturbulent velocity ($\xi_t$) have been obtained. The metallicity is also one of the fundamental
parameters of the atmosphere, as it controls the opacity in the continuum. It is obtained only by an
iterative process, and is not limited to \FeH but includes the abundance of all electron suppliers, such as the
$\alpha$--elements Mg or Si. We plan to include the ratio of the $\alpha$--elements to iron, [$\alpha$/Fe],
in the next edition of the catalogue, both because it is an important parameter for a better
characterization of the metallicities, but also because it is an important parameter in population studies.

The present version of the catalogue has been built up from the previous one (Cayrel de Strobel et al 
\cite{cay97}), which was complete up to December 1995,
keeping 258 references, corresponding to
3880 determinations of atmospheric parameters for 2484 stars. 
120 references from refereed journals, published  
between January 1996 and  December 2000, have been added to the catalogue.
They correspond to 2474 new determinations, and 873 new analysed stars.  

 The presentation of the catalogue, completely revised and reformatted, is
described in Sect. 2. Some comments about the input data, the stellar 
content of the catalogue and its connection to the Hipparcos mission are given in
Sect. 3. Concluding remarks are given in Sect. 4.


\section{Description of the catalogue}

The format of the catalogue was modified compared to the previous version 
in order to make its use more convenient. When possible, the field stars have been identified 
by three designations to make the cross-reference
between papers easier. 
The basic data on stars include ICRS 2000.0 coordinates, apparent V magnitude and
spectral type. The atmospheric parameters are given with their 
error bars when available. \\

As usual \FeH is defined by :

[Fe/H] = $\log$(Fe/H)$_{\rm star} - \log$(Fe/H)$_{\rm Sun}$\\

where Fe/H is the ratio of the number of iron atomes to the number of hydrogen atomes
in the atmosphere of either the star and the Sun.
Values in the previous versions given with respect to a standard star other than
the Sun have been 
converted to the solar scale.
Another improvement of the catalogue concerns the references which are no longer given in a 
separate table, but in the last column of Tables 1 and 2. They are presented in the form of 
standard CDS and ADS codes which allow a direct
access to the publication.\\

Tables 1 and 2, corresponding respectively to field stars and to stars in clusters or external galaxies,
 contain the following columns : \\

\begin{enumerate}

\item {\it Identifiers}\\
For field stars, 3 identifiers are proposed. The first column presents an identifier
which was chosen according to the following rule : HD is chosen preferentially if available, if
not available, BD is chosen, then CD/CPD, then Giclas. Such a rule allowed us to gather together
all the \FeH \, determinations for the same star.  The HIP
number is also given for more than 90\% of the stars in Table 1, as well as an
alternate designation for the bright stars which are often designated in the literature 
by their name or their number in a constellation. 
 For cluster 
stars, a great variety of
names were found in the literature for the same object. 
The SIMBAD database was consulted 
in order to adopt the most appropriate designation for each star. Except for
51 objects not recognized by SIMBAD (designated with "?" at the beginning of their name), the chosen
identifier can be used in a SIMBAD interrogation. In many cases, problems related to identifiers were
solved thanks to the WEBDA\footnote{http://obswww.unige.ch/webda/} database  devoted to stellar open clusters.
 There is no
redundancy between the list of field stars and the list of cluster stars. We encourage authors of 
abundance analyses to verify the syntax of the identifiers they use in the SIMBAD and WEBDA databases 
or the Dictionary of Nomenclature 
of Celestial Objects (Lortet et al. \cite{lor94}) and to use 
preferentially the HD number for field stars.\\

\item {\it ICRS 2000.0 equatorial coordinates}\\
The ICRS 2000.0 equatorial coordinates have been collected through SIMBAD or WEBDA, but they are not 
available for 
all the stars. They have been included in the catalogue for optimal use of the VizieR Service (Ochsenbein et 
al \cite{och00}).\\

\item {\it Visual magnitude V}\\
The visual V magnitudes are from the SIMBAD database. The sources of the visual magnitudes 
in SIMBAD 
are various and heterogeneous and as a 
consequence, this value should be considered only as an indicator of
brightness. For precise photometry, the users have to consult
specialised catalogues which are included in the General Catalogue of Photometric 
Data\footnote{http://obswww.unige.ch/gcpd/gcpd.html}
(Mermilliod et al \cite{mer97}). In some cases, the magnitude indicated in this column is
the B magnitude (a letter B follows the value of the magnitude in this case). 
Only a few faint stars do not have any 
visual magnitude at all.\\

\item {\it Spectral type}\\
The spectral types come from the cross-identification of the catalogue and 
the SIMBAD database. The same syntax has been used (see the SIMBAD user's guide 
and reference manual, chapter 15). We have corrected some spectral types which 
were clearly in disagreement with the effective temperature and gravity resulting from a 
detailed analysis, especially for metal deficient population II stars. 
In particular, for a fairly numerous  sample of population II bright yellow
giants, we found it necessary to give a more
advanced spectral type
and a  brighter luminosity class, reflecting more correctly their
position in the HR diagram. This misclassification of very evolved population II
stars, still present in  major stellar catalogues, is due
to the great metal deficiency  of their atmosphere. The MK spectrum of
such stars mimics the MK spectrum of a  hotter unevolved star.  To such stars
we  have  assigned a more appropriate spectral type, KIIvw (the "vw" stands for 
"very weak" lines).\\

\item {\it Effective temperature \Teff and its error}\\
The value listed is the one which was adopted by the author for the abundance 
determination in the detailed analysis. The effective temperature of a star, which is the 
critical parameter, is mainly derived from narrow-band photometry, or on 
purely spectroscopic grounds from the comparison between $H_{\alpha}$
observed profiles and $H_{\alpha}$ computed profiles. 

When available, the error on \Teff determined by the author is also given. The best 
determinations 
of \Teff listed in the catalogue quote errors of 25 K, whereas they can reach 250 K 
in some cases (faint
stars, cold stars, unresolved stars...).
\\

\item {\it Logarithm of gravity \logg and its error}\\
The value of \logg is the one used by
the authors in the spectrum analysis. Usually, the spectroscopic surface gravity
is determined from ionisation and excitation equilibria, as obtained from neutral and
ionized lines, carefully chosen in the stellar spectrum, and from wings of strong lines, 
broadened
by collisional damping. In some recent papers, the Hipparcos parallax was used to determine 
the gravity. The error on \logg is given if available.\\

\item {\it \FeH and its error}\\
 Contrary to the previous versions of the catalogue, \FeH is always given with respect to 
the Sun.
Values in the previous versions given with respect to another standard star have been converted 
to the
solar scale. It is worth noticing that the solar scale can change from author to author, 
with $\log \epsilon_\odot(Fe)$ varying from 7.47 to 7.67. Several determinations have been flagged, by the letter M
when [M/H] is given instead of \FeH and by the letter N when the Fe abundances is based on NLTE analysis 
(Th\'evenin \cite{the99}). The error on \FeH is given if available.\\

\item {\it Reference}\\
The reference of each \FeH determination includes the name of the first author in an abbreviated form and 
the standard 
reference code (bibcode) of the paper. In such way, all the papers quoted in the catalogue can be retreived easily
through ADS or VizieR Service. Only the following journals have been searched for
\FeH determinations : A\&A, A\&AS, AJ, ApJ, PASP, PASJ, NewA, MNRAS.

   \end{enumerate}

As in the last two editions of the catalogue (1991, 1997), the microturbulence velocity $\xi_t$ has 
been omitted due to the fact that not all
the authors use the same definition for it. This parameter must be recovered from the 
original reference.

The field stars list (Table 1) includes 4918 determinations of [Fe/H] for 2447 different stars. The second part of
the catalogue (Table 2) includes 1436
determinations of [Fe/H] for 909 stars in 79 stellar systems.

 We have been requested several times, between the successive editions, to include 
an average of the different determinations of [Fe/H] for each star. This is outside the scope of the [Fe/H]
catalogue, which is purely bibliographical, but a paper presenting averaged atmospheric parameters for
a sub-sample of the catalogue is under preparation.


\section{Some comments on the catalogue}

\subsection{Input data}

It is interesting to have a look at the growth of the catalogue over 20 years. Fig. 1 
presents the evolution per year of the number of \FeH determinations, 
the number of new stars included in the catalogue and the number of papers presenting 
\FeH determinations 
for FGK stars. The peak in 1990 is mainly due to the papers of McWilliam \cite{mcw90} 
(668 determinations) and
Balachandran \cite{bal90} (189 determinations). The number of papers has been growing slowly 
up to 2000, as 
has the number of new stars which underwent a detailed analysis. The number of 
\FeH has been increasing
faster, especially during the year 2000. As a matter of fact, more and more often the determination 
of \FeH is only the first 
step in studying other elements in stars which
are already known to belong to a given population. For this reason, some 
stars might be analysed several times by the same author or by different authors interested in elements
other than iron.  Also, 
a few stars are used as comparison stars to test a method and have many determinations of atmospheric
parameters. As an example, the star which is the most studied is the metal poor halo subgiant HD 140283
(30 \FeH determinations between 1980 and 2000).
There is more than 400 K difference between the temperature proposed by Magain \cite{mag84} (5419
K) and that of Fuhrmann et al \cite{fuh97} (5843 K). The iron abundance \FeH consequently has a large 
range of values,
from -3.06  (Magain \cite{mag84}) and -2.21 (NLTE, Th\'evenin \cite{the99}) or -2.29 (Zhao \cite{zha00}, 
Fuhrmann \cite{fuh98}). This shows that even with high 
quality observations and 
a careful analysis, the atmospheric parameters vary from author to author.


\begin{figure}[t]
\resizebox{\hsize}{!}{\includegraphics{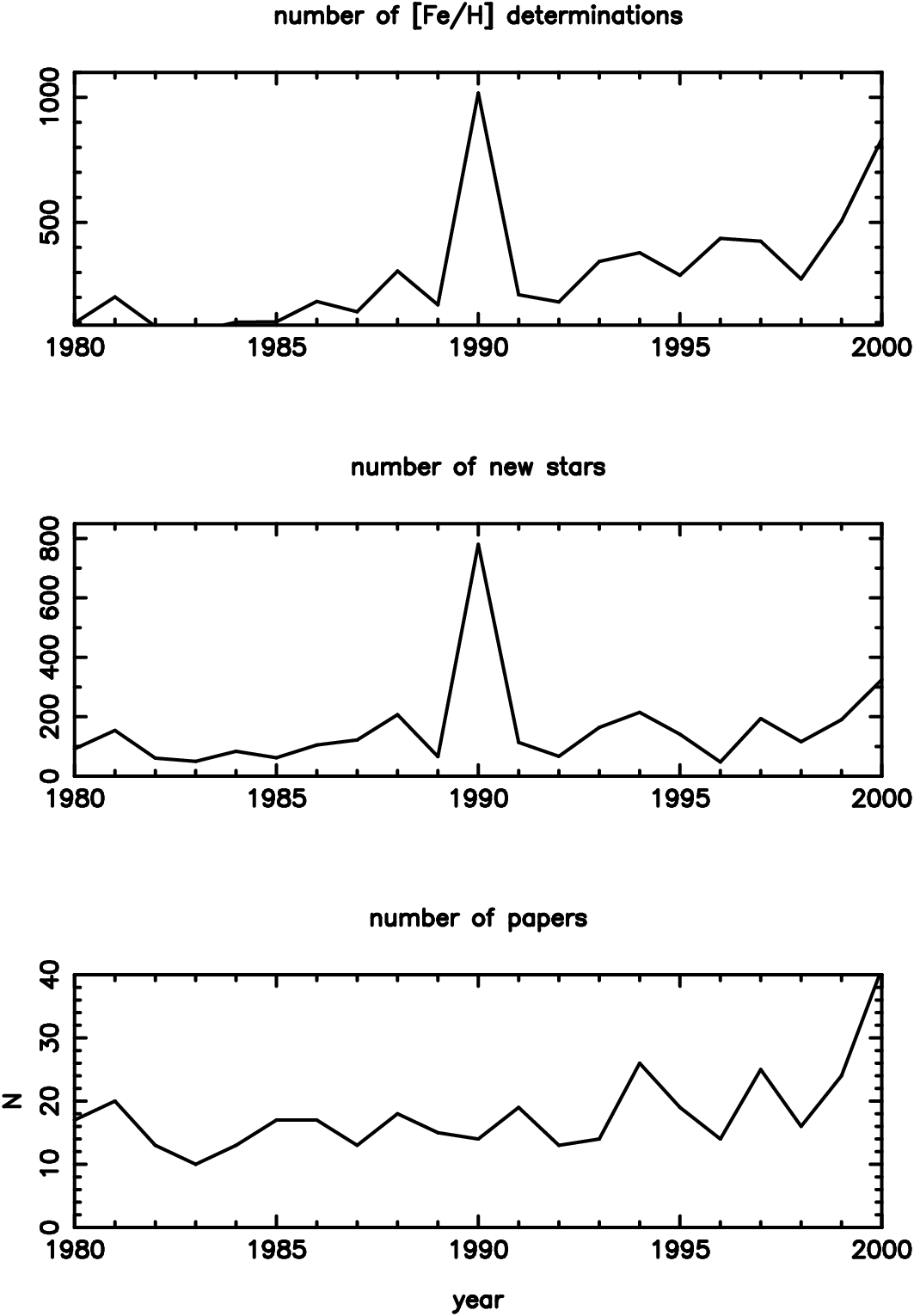}}
\caption{From top to bottom, evolution of the number per year of \FeH determinations, of new stars 
introduced in the catalogue and of papers quoting spectroscopic \FeH determinations for FGK stars.}
\end{figure}

\subsection{Stellar content of the catalogue}

The great change of this new edition of the catalogue of \FeH \, determinations is the restriction to 
middle and low mass FGK stars which span a large range of ages. In its present form,  the catalogue 
is principally suited to studies of the chemical evolution of the Galaxy. The distribution of the 6354
\FeH \, determinations of the catalogue is shown in the plane (\Teff, \logg) in Fig. 2, in the 
plane (\Teff, \FeH) in Fig. 3 and in the histogram of Fig. 4. In Fig. 2, the Herzsprung-Russell gap, due to
the extremely rapid evolution of the stars in their subgiant phase, is visible
between \logg $\sim$ 3.0 and \logg $\sim$ 4.2. In Fig. 3, the separation between extreme Population II 
subdwarfs (\Teff $\sim$ 6000 K) and extreme Population II bright giants (\Teff $<$ 4800 K) is clearly seen.
This is mainly an observational selection effect, extremely metal-poor dwarfs being picked-up by surveys only
if they are bright enough (i.e. near the turn-off).

Despite the improvement of telescopes and spectrographs, there is still  a lack of 
G and K dwarfs, which are intrinsically faint and more difficult to observe at high resolution 
and high S/N 
than the giants corresponding  to the same \Teff.  A few M stars have been introduced in the
catalogue but they are largely underepresented because 
they are difficult to analyse in detail.
 
The sample of stars in this new edition of the catalogue, with \Teff $<$ 7000 K,
cannot be considered as
representative of the stellar content of the solar neighbourhood. 
 Evidently, the various observing programs, spanning from stellar structure problems to stellar 
population studies, from 
which the catalogue 
was built, introduce some biases in the distributions of \tgm.

\begin{figure}[t]
\resizebox{\hsize}{!}{\includegraphics{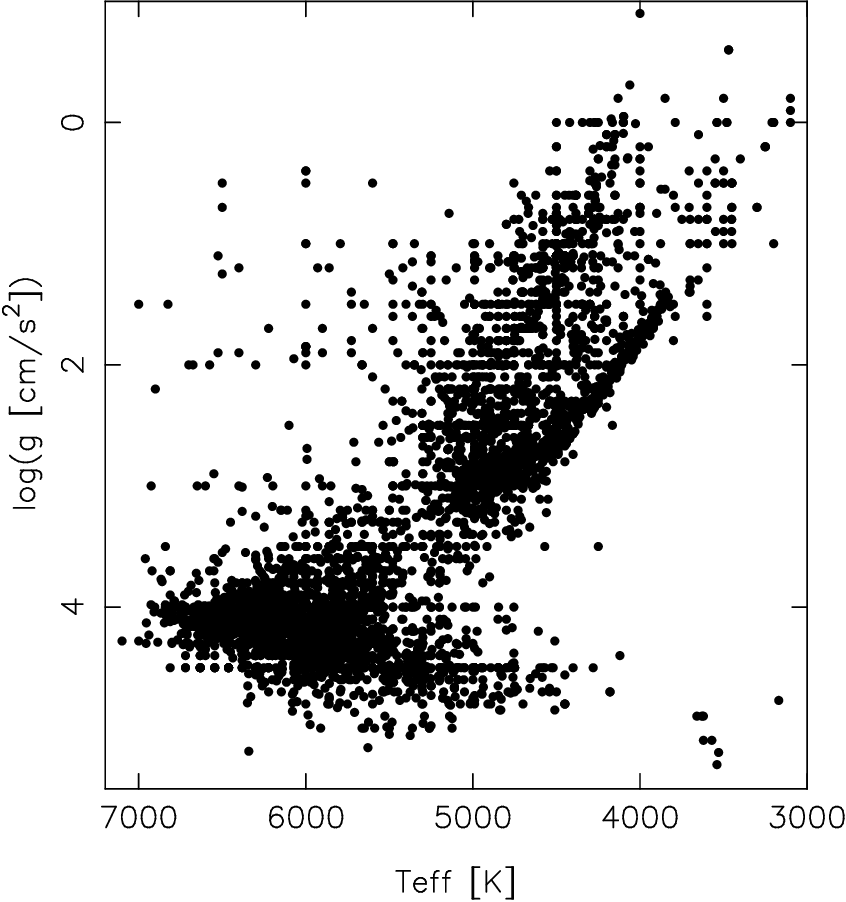}}
\caption{\Teff vs. \logg for the 6354 entries of the catalogue.}
\label{Teff_logg}
\end{figure}


\begin{figure}[t]
\resizebox{\hsize}{!}{\includegraphics{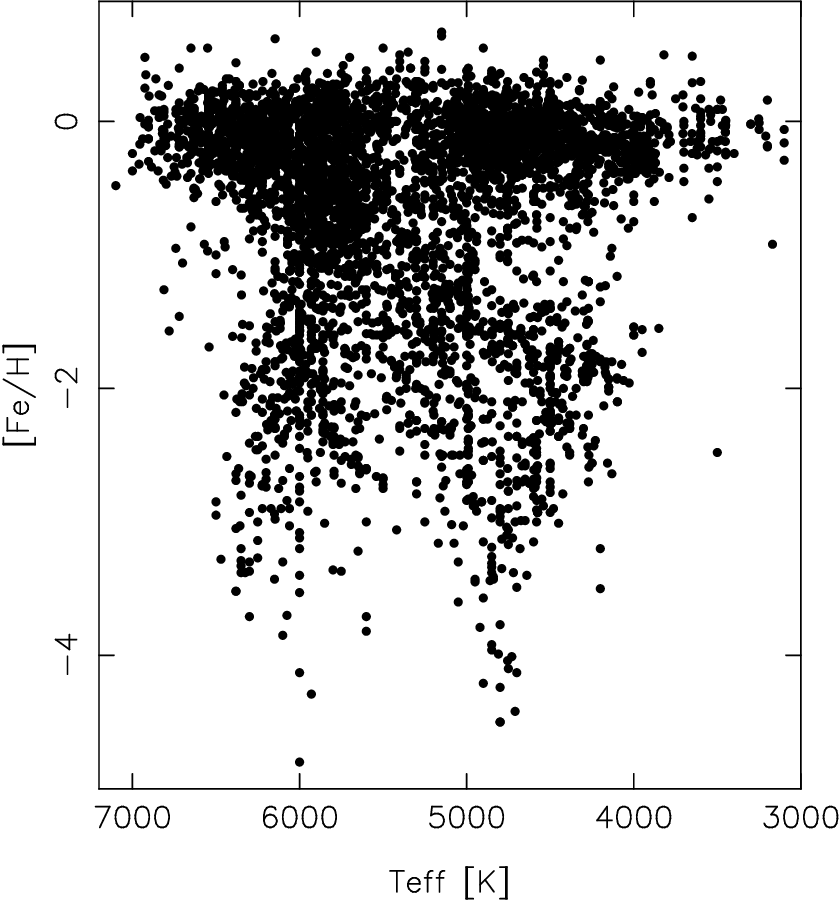}}
\caption{\Teff vs. \FeH for the 6354 entries of the catalogue.}
\label{Teff_FeH}
\end{figure}


\begin{figure}[t]
\resizebox{\hsize}{!}{\includegraphics{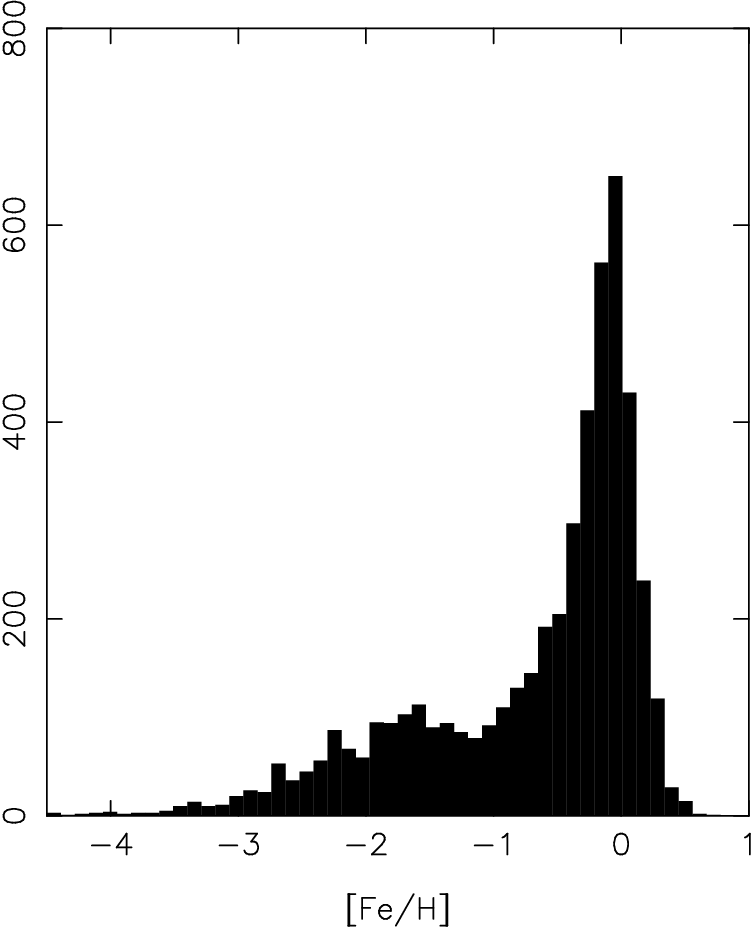}}
\caption{\FeH histogram for the 6354 entries of the catalogue.}
\label{FeH_histo}
\end{figure}


\subsection{The Catalogue and the Hipparcos Mission}

Many stars contained in this catalogue were included in different programs
linked to the Hipparcos mission. The crossing between the Hipparcos data
and the spectroscopic results gathered in this catalogue has a strong 
impact on the methods of analysis of stellar spectra.  In particular, 
thanks to Hipparcos, spectroscopic gravities, based on ionisation
equilibrium, have shown to be in error in very metal poor stars 
(Nissen \cite{nis97}, Fuhrmann \cite{fuh98}). The  Hipparcos number of the stars is
given,  when available, in column 2 of Table 1.

\section{Conclusion}

We have presented the new version of the catalogue of [Fe/H] determinations
which is now restricted from middle to low  mass F, G and K stars.
The [Fe/H] values contained in  the catalogue come almost all  from
differential detailed analyses. Greenstein introduced, in the late
fifties, the technique of the differential 
curve of growth analysis. The principle of 
 differential detailed analyses  is to obtain the abundances of the
elements 
in star A relative to the abundances in star B, taken as a standard. 
If we take the Sun as a standard, and if we restrict the effective
temperature interval to F, G, and K stars, the same spectral lines can
be used, and the knowledge of the oscillator strengths is no longer
needed. Another advantage of this method is that it cancels systematic
errors in equivalent width measurements if the same spectroscopic equipment
is used to get the spectra of both stars. A third advantage of the
differential method 
is that the effects of departure from local thermodynamic equilibrium (LTE) are
minimized because they are expected to be about the same in both stars.
Nevertheless, if we take a quick  look at  
the catalogue, concentrating on stars analysed several times  
in  Table 1,  we see 
that the differences between some  authors are still, in the mean, 
higher than the standard errors attributed to each analysis. Let us
hope that these differences will be minimized with the future progress both in 
observation and theory.

\begin{acknowledgements}
Our thanks go to the successive collaborators of the catalogue, who 
spent energy and time working for it. We are also very thankful to those
of our colleagues who provided their data in electronic form.
Many thanks  to Monique and Fran\c cois Spite for sending remarks and corrections over many 
years. We thank also Jean-Louis Halbwachs and Fran\c cois Ochsenbein for 
pointing out some errors in the catalogue, and the referee, J.-C. Mermilliod, for valuable comments.
We made extensive use of the CDS-SIMBAD and NASA-ADS databases and VizieR Service at CDS, and we are 
extremely grateful to the staff of these services for maintaining such
valuable resources and for their assistance.
\end{acknowledgements}

\end{document}